\documentclass[aps,groupaddress,superscriptaddress,reprint,amsmath,amssymb,graphicx]{revtex4-1}
\usepackage[utf8]{inputenc}
\usepackage{natbib}
\usepackage{color}
\usepackage{xcolor}
\usepackage{graphicx}
\usepackage{mdframed}
\usepackage{physics}
\usepackage{bm}

\begin{document}
\preprint{APS/123-QED}
\title{Information-driven transitions in projections of underdamped dynamics}
\author{Giorgio Nicoletti}
\affiliation{Department of Physics and Astronomy ``G. Galilei'', University of Padova, Padova, Italy}
\author{Amos Maritan}
\affiliation{Department of Physics and Astronomy ``G. Galilei'', University of Padova, Padova, Italy}
\author{Daniel Maria Busiello}
\affiliation{Institute of Physics, \'Ecole Polytechnique F\'ed\'erale de Lausanne - EPFL, 1015 Lausanne, Switzerland}

\begin{abstract}
\noindent Low-dimensional representations of underdamped systems often provide insightful grasps and analytical tractability. Here, we build such representations via information projections, obtaining an optimal model that captures the most information on observed spatial trajectories. We show that, in paradigmatic systems, the minimization of the information loss drives the appearance of a discontinuous transition in the optimal model parameters. Our results raise serious warnings for general inference approaches and unravel fundamental properties of effective dynamical representations, impacting several fields, from biophysics to dimensionality reduction.
\end{abstract}

\maketitle

Data-driven approaches to infer dynamical features from single trajectories are solidly taking hold, especially as the spatiotemporal resolution of experimental data is rapidly increasing. This inference problem is well understood for deterministic systems \cite{crutchfield1987equations, daniels2015automated, brunton2016discovering}, while criticalities arise in stochastic systems, where fast variables have to be treated as external noise \cite{risken1996fokker}. Although fundamental progresses were recently made for stationary underdamped stochastic processes \cite{bruckner2020inferring}, most of the recent studies focused on stationary overdamped systems \cite{el2015inferencemap, garcia2018high, frishman2020learning, ragwitz2001indispensable,gnesotto2020learning}, where the fast equilibration of velocities is employed. Crucially, it has been shown that employing this simplification \textit{ab initio} might lead to erroneous results \cite{liang2021intrinsic,lau2007state}, whose origin dates back to the Ito-Stratonovich dilemma \cite{kupferman2004ito}. Nevertheless, the overdamped framework remains a paramount tool to gain analytical insights.

Yet, the problem of reducing the dynamics in the full position-velocity phase-space, $(\vec{x},\vec{v})$, to an effective dynamics in the $\vec{x}$-space is challenging and far-reaching. Indeed, it might impact different fields, ranging from coarse-graining procedures \cite{katsoulakis2003coarse,gfeller2007spectral,altaner2012fluctuation} to dimensionality reduction techniques and machine learning approaches \cite{coifman2008diffusion, wehmeyer2018time, otto2019linearly, swischuk2019projection, wright2022high}. Here, we propose a method to derive effective dynamics that require no prior knowledge of the system.

As information theory has proven to be a promising framework to capture essential features of complex stochastic systems \cite{tkavcik2016information,nicoletti2021mutual, mariani2021critical}, we build an optimal model that captures the maximum amount of information on possibly short-time trajectories in the $\vec{x}$-space. Hence, we relax the stationarity assumption and include the effect of the initial conditions. We show that the information-preserving feature of our approach is associated with an unforeseen discontinuity in the parameter space of the optimal model. This result poses a significant and sweeping limitation on the efficacy of effective models in predicting underlying dynamics, as slight changes of underdamped parameters may give rise to large variations in the optimal prediction. Notwithstanding, the proposed method has a broad applicability, e.g., passive tracers in active media \cite{dabelow2019irreversibility}, species dynamics in ecological communities \cite{vacher2016learning}, effective models to probe neural activity \cite{mastrogiuseppe2018linking, williams2018unsupervised}, or any dynamics with unobserved degrees of freedom.

\begin{figure}[t]
    \centering
    \includegraphics[width=\columnwidth]{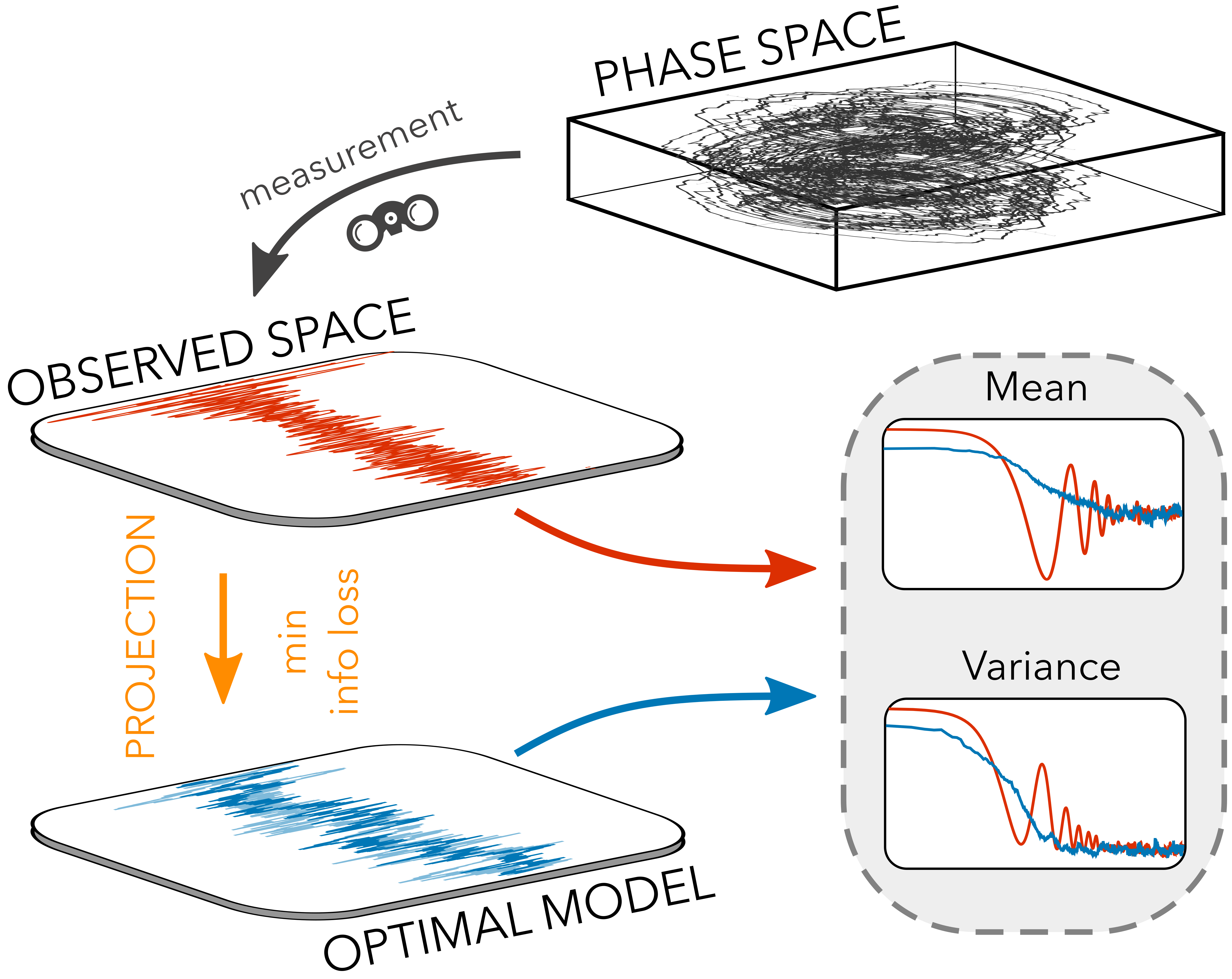}
    \caption{Sketch of the method. The observed trajectories reflects an underlying dynamics in a higher-dimensional phase space. We build an optimal model capturing the observed evolution via a projection that minimizes the information loss. Analytically solvable optimal models will capture at best essential observed features, such as mean and variance.}
    \label{fig:figure1}
\end{figure}

Consider a system whose dynamics is described by an underdamped model,
\begin{eqnarray}
    \label{under}
    \dot{\vec{x}} &=& \vec{v} \\
    \dot{\vec{v}} &=& - \gamma \vec{v} + \vec{F}(\vec{x}) + \sqrt{2} \hat{\Delta} \vec{\xi}(t) \nonumber
\end{eqnarray}
where $\gamma$ is the friction coefficient, $\hat{\Delta}^T \hat{\Delta} = \hat{D}$ is the diffusion matrix, and $\vec{F}$ a generic non-linear position-dependent force. Mass is set to $1$ for simplicity. This general model also includes chiral diffusion \cite{hargus2021odd}. However, it is often the case that the details of Eq.~\eqref{under} are not known, or the model cannot be solved analytically.

\begin{figure*}[t]
    \centering
    \includegraphics[width=\textwidth]{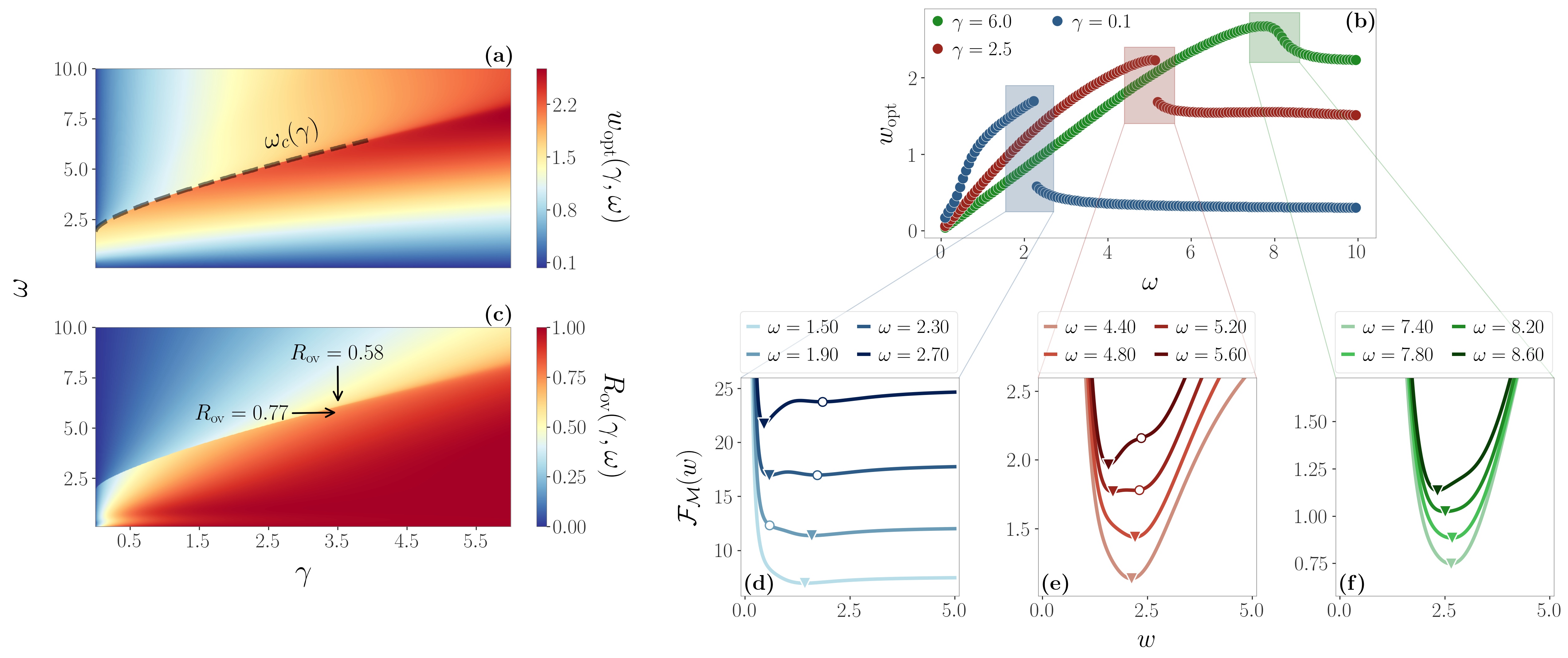}
    \caption{(a) Contour plot of $w_{\rm opt}(\gamma, \omega)$ in the harmonic case, showing a discontinuity line $\omega_c(\gamma)$. (b) By plotting $w_{\rm opt}(\omega)$ for selected values of $\gamma$, we see that the discontinuity gradually decreases as $\gamma$ increases and eventually disappears. At large $\omega$, $w_{\rm opt}$ displays a plateau. (c) The ratio $R_{\rm ov}(\gamma, \omega)$ shows that for high values of $\gamma$ our method coincides with the overdamped limit. When the transition line is crossed, we find drastically different values of $w_{\rm opt}$ even for relatively large $\gamma$. (d-f) The transition is driven by the presence of two minima of $\mathcal{F}_\mathcal{M}$ that exchange stability. The global minimum is highlighted by a triangle, the unstable minimum (if present) by a circle. At low enough values of $\gamma$, as we increase $\omega$ a second minimum appears at low $w$. These two minima eventually coalesce at larger $\gamma$, smoothing the transition.}
    \label{fig:figure2}
\end{figure*}

Therefore, we seek a method to build a solvable effective model in lower dimensions. In particular, it is often the case that we only have access to short-time trajectories in the $\vec x$-space. Building on these premises, our method declines in two steps, sketched in Fig.~\ref{fig:figure1}. First, we estimate the probability density function (pdf) at all times in the $\vec x$-space, $P_{\rm exp}({\vec x},t)$, which ideally coincides with the exact solution of Eq.~\ref{under} marginalized over the $\vec v$-space. Then, we introduce an information projection that maps this pdf into the solution of an optimal effective model, minimizing the information loss.

In principle, $P_{\rm exp}({\vec x},t)$ should be estimated from data, but this task is often unfeasible. Here, we only consider the experimental mean, $\vec{\mu}_{\rm exp}(t)$, and covariance matrix, $\hat{\sigma}_{\rm exp}(t)$. Hence, the simplest unbiased description for the marginal pdf is a multivariate Gaussian, $P_{\rm exp}(\vec{x},t) = \mathcal{N}(\vec{\mu}_{\rm exp}(t), \hat{\sigma}_{\rm exp}(t))$. This corresponds to a maximum entropy ansatz \cite{jaynes2003probability} at each time. Notice that $\vec{\mu}_{\rm exp}(t)$ and $\hat{\sigma}_{\rm exp}(t)$ depend also on $\gamma$, $\vec{F}$ and $\hat{\Delta}$ in non-trivial ways, and in general there are no stochastic Markov processes whose pdf coincides with the proposed form of $P_{\rm exp}(\vec{x},t)$. 

An effective model with homogeneous coefficients retaining the Gaussian form of $P_{\rm exp}(\vec{x},t)$ is an Ornstein-Uhlenbeck (OU) process of the following form:
\begin{equation}
    \dot{\vec{x}}(t) = - w_p \left(\vec{x}(t) - \vec{\mu}_p \right) + \sqrt{2} \hat{\Delta}_p \vec{\xi}_p(t),
    \label{OU}
\end{equation}
where $\vec{\xi}_p$ is a white noise, and $\{ w_p$, $\vec{\mu}_p$, $\hat{\Delta}_p \} := {\vec \theta_p}$. We choose these parameters so that the information loss between $P_{\rm exp}$ and the solution of Eq.~\eqref{OU}, $P_{\rm OU}$, is minimal over the entire trajectory duration, $t_{\rm exp}$:
\begin{equation}
    {\vec \theta_{\rm opt}} = \underset{{\vec \theta_p}}{\rm argmin} \int_0^{t_{\rm exp}} dt~\mathcal{M}(P_{\rm exp}, P_{\rm OU})
    \label{minM}
\end{equation}
where $\mathcal{M}$ is any information metric. Eq.~\eqref{minM} defines our information projection. The pdf of the corresponding optimal model is $P_{\rm opt}(\vec{x}, t) = P_{\rm OU}(\vec{x}, t \, | \,\vec{\theta}_{\rm opt})$. To avoid divergences in the stationary limit, we ask that $\mathcal{M} \to 0$ for $t_{\rm exp} \to +\infty$, resulting in a constraint on the stationary mean and variance of $P_{\rm opt}(\vec{x},t)$. However, the method can still be applied without this constraint, as shown in the Supplemental Material \cite{supplemental_material}. We remark that, in principle, it is possible to apply this approach starting from any $P_{\rm exp}$, estimated from data, and building a information projection into any desired model, eventually losing analytical tractability.

As a proof of concept, we apply the method to systems with one spatial dimension. The underdamped dynamics lives in a $2D$ phase space and the optimal model is a $1D$ OU process. The multidimensional extension is conceptually straightforward, but deserves proper attention in dealing with non-diagonal diffusivities.

First, we consider the paradigmatic case of an harmonically-bounded particle, i.e., in Eq.~\eqref{under} we set $F = \omega^2 x$ and $\hat{\Delta} = \sqrt{D} = \sqrt{k_B T\gamma^2}$ for thermodynamic consistency \cite{risken1996fokker}. With this choice, Eq.~\eqref{under} can be solved exactly, and its propagator is Gaussian. For Gaussian initial distributions of $x$ and $v$, $\mathcal{N}_{x_0}(\mu_{x_0},\sigma_{x_0})$ and $\mathcal{N}_{v_0}(\mu_{v_0},\sigma_{v_0})$, the resulting $P_{\rm exp}(x, t)$ is Gaussian as well, and the maximum entropy ansatz is exact. In this scenario, the analytical expressions of $\mu_{\rm exp}(t)$ and $\sigma_{\rm exp}(t)$, along with the $t_{\rm exp} \to +\infty$ limit, allow us to single out the properties of the optimal model defined by the second step of our method, Eq.~\eqref{minM}. Since the steady state of $P_{\rm OU}$ is fixed, the effective model, Eq.~\ref{OU}, is fully specified by one parameter, e.g., ${\vec \theta}_p = \{w_p\}$ (see Supplemental Material \cite{supplemental_material}).

In Figs.~\ref{fig:figure2}a-b we show that the resulting $w_{\rm opt}(\gamma, \omega)$ exhibits a line of discontinuities at $\omega = \omega_c(\gamma)$. These plots are obtained for $\mathcal{M} = D^{\rm sym}_{\rm KL}$, which is the symmetrized Kullback-Leibler divergence \cite{ThomasCover2006}. In the Supplemental Material \cite{supplemental_material}, we show that this discontinuity is affected by the initial conditions and survives for different choices of $\mathcal{M}$, e.g., the Hellinger distance, the geodesic distance, the Chernoff-alpha divergence, and the Wasserstein distance \cite{ThomasCover2006, amari2016information}. We can also compare the values of $w_{\rm opt}$ with the ones predicted by an overdamped limit, $w_{\rm ov} = \omega/\sqrt{\gamma}$, a classical projection in the $x$-space usually employed for strong friction regimes. By plotting the ratio $R_{\rm ov} = w_{\rm opt}/w_{\rm ov}$, it is evident that the optimal model is markedly different from the overdamped model, even at relatively large values of $\gamma$ (see Fig.~\ref{fig:figure2}c).

\begin{figure}[t]
    \centering
    \includegraphics[width=\columnwidth]{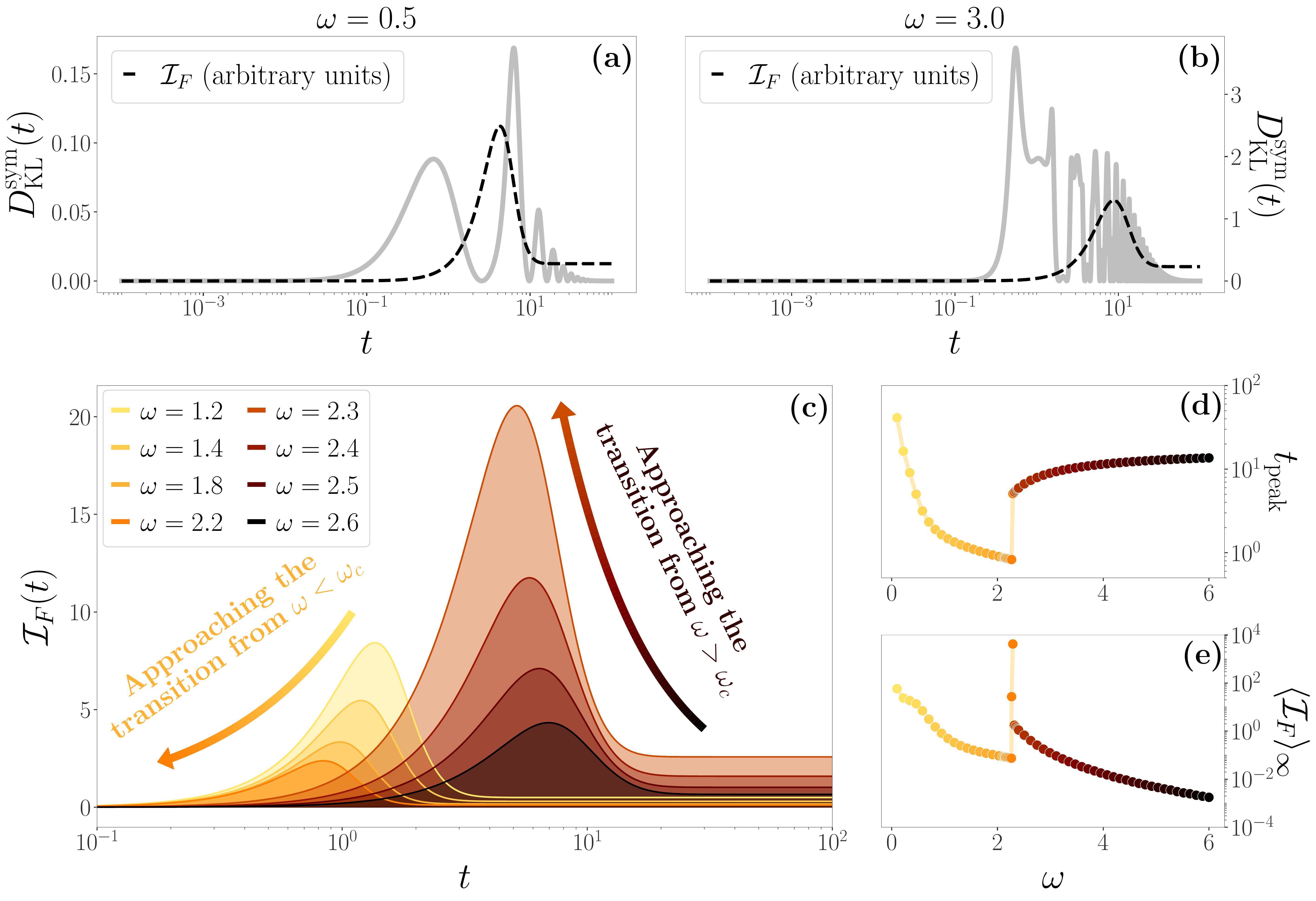}
    \caption{(a-b) Plot of $D^{\rm sym}_{KL}(t)$ (gray) and $\mathcal{I}_F(t)$ (black) in the harmonic case before and after the transition, respectively. (c) $\mathcal{I}_F(t)$ exhibits opposing trends before and after the transition. (d) These trends are well characterized by the peak time of $\mathcal{I}_F$, $t_{\rm peak}$, which captures the transient dynamics for $\omega < \omega_c$, and the oscillatory dynamics for $\omega > \omega_c$. Notice that at small $\omega$ we are close to the free diffusion regime with a longer transient. (e) At the transition, the integral mean of the Fisher information diverges. In all these plots, $\gamma = 0.1$, $\sigma_{x_0}^2 =  \sigma_{v_0}^2 = \mu_{x_0} = v_{\rm th} = 1$ and $\mu_{v_0} = 0$.}
    \label{fig:figure3}
\end{figure}

The integral of the information metric, $\mathcal{F}_\mathcal{M} = \int  dt\, \mathcal{M}$, plays a role similar to a free energy, whose global minimum defines the optimal model. The discontinuity can then be seen as a first-order phase transition due to the presence of two minima that exchange stability (see Figs.~\ref{fig:figure2}d-f). To investigate the meaning of the phases associated with these minima, we introduce the Fisher information \cite{ThomasCover2006, amari2016information}.
\begin{gather}
    \mathcal{I}_F(\omega,t \,|\,\gamma) = \int_{-\infty}^{+\infty} \frac{\partial \log(P_{\rm opt}(x, t \, | \, w_{\rm opt}(\omega,\gamma)))}{\partial \omega} \bigg|_\gamma dx \nonumber \\
    = \frac{(\partial_\omega w_{\rm opt})^2\left[2 \sigma_{\rm opt} (\partial_{w_{\rm opt}} \mu_{\rm opt})^2 + (\partial_{w_{\rm opt}} \sigma_{\rm opt})^2\right]}{2 \sigma_{\rm opt}^2}\bigg|_\gamma
    \label{fisher}
\end{gather}
where $\mu_{\rm opt}$ and $\sigma^2_{\rm opt}$ are the mean and variance of the optimal pdf, respectively, and $\partial_{w_{\rm opt}} = \partial / \partial w_{\rm opt}$. Eq.~\eqref{fisher} quantifies the sensibility of $P_{\rm opt}$ to changes in $\omega$, at a fixed value of $\gamma$. In Fig.~\ref{fig:figure3}a-b we show the temporal evolution of $\mathcal{I}_F$. Approaching the transition from below ($\omega \lesssim \omega_c$), $\mathcal{I}_F$ peaks at short times, indicating that the information projection weights more earlier stages of the dynamics, i.e., the transient regime. On the other hand, for $\omega \gtrsim \omega_c$, the peak of $\mathcal{I}_F$ appears at longer times, capturing the persistent oscillating behavior. Notice that, for very small values of $\omega$, the system is close to the free-diffusion regime. This reflects into longer transients and, in turn, an increase of the peak time (see Figs.~\ref{fig:figure3}c-d). The integral mean $\ev{\mathcal{I}_F}_T = \frac{1}{T} \int_0^T dt \, \mathcal{I}_F$ in the limit $T\to+\infty$ quantifies the total susceptibility of the optimal model to changes in $\omega$. This quantity diverges at the transition point, as expected. Moreover, we observe that $\ev{\mathcal{I}_F}_\infty$ considerably decreases at large values of $\omega$ since $w_{\rm opt}(\omega)$ saturates, indicating an increasing robustness of the information projection (see Fig.~\ref{fig:figure3}e).

These results suggest that the two phases of the optimal model, represented by the minima of $\mathcal{F}_\mathcal{M}$, are characterized by the dynamical regimes they capture the most, i.e., transient or persistent oscillations. Crucially, in the underdamped model the dynamics changes smoothly across the transition line, highlighting that the information-preserving feature of the projection is at the root of the discontinuous transition. Remarkably, as shown in the Supplemental Material \cite{supplemental_material}, it does not appear if $\mathcal{M}$ is not an information metric, e.g., the $L^2$-distance between the mean and variance of $P_{\rm exp}$ and $P_{\rm OU}$.

Although in general there is no analytical expression for $w_{\rm opt}$, in some regions of model parameters it is possible to gain an analytical grasp of its form. By tuning the system so that the variance stays constant at all times, i.e., $\sigma_{x_0}^2 = k_B T/\omega^2$ and $\sigma_{v_0}^2 = k_B T$, the Kullback-Leibler divergence depends only on the mean. Eq.~\eqref{minM} can be solved analytically for $\mu_{v_0} = 0$, and the solution can be expanded as follows:
\begin{gather}
w_{\rm opt} \approx \frac{\omega}{\sqrt{\gamma}}\left[1 - \frac{3}{8}\left(\frac{\omega}{\gamma}\right)^4\right] \qquad \gamma \to +\infty \\
w_{\rm opt} \approx \sqrt{\omega} \left[\left(1+\frac{2}{\sqrt{3}}\right)^{1/4} - \frac{(3+2\sqrt{3})^{1/4}}{2\sqrt{6}\omega}\gamma\right] \qquad \gamma \to 0 \nonumber
\end{gather}
at the leading orders. The large $\gamma$ limit shows the next-to-leading order corrections to the overdamped expression, $\omega_{\rm ov}$, while the small $\gamma$ regime unveils a drastically different scaling. In this case, the exact solution exhibits no transition, suggesting that the two phases of the effective model emerge from an interplay between the optimization of the time evolution of both mean and variance.

\begin{figure*}[t]
    \centering
    \includegraphics[width=\textwidth]{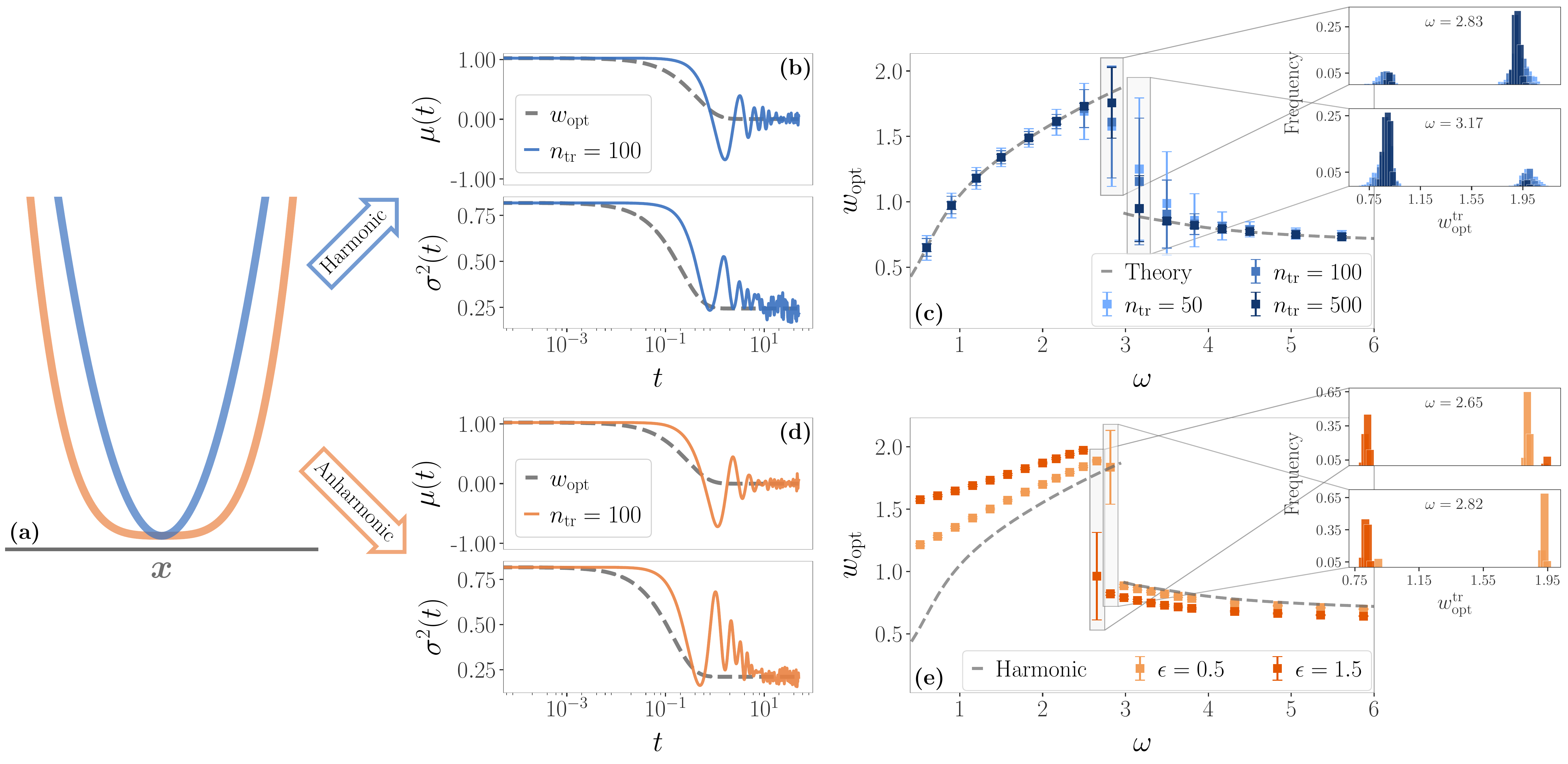}
    \caption{Results obtained from a limited number of spatial trajectories, in the presence of harmonic and anharmonic potentials. For all these plots, $\sigma_{x_0}^2 =  \sigma_{v_0}^2 = \mu_{x_0} = v_{\rm th} = 1$, $\mu_{v_0} = 0$ and $\gamma = 0.1$. (a) Depiction of the two potentials. (b) Mean and variance in the harmonic case, obtained from $n_{\tr} = 100$ trajectories. The gray dashed line represents the mean and variance of the corresponding optimal model. (c) The optimal value $w_{\rm opt}$ in an harmonic potential, as a function of $\omega$ and for different numbers of trajectories. The error bars represent one standard deviation over $10^3$ numerical experiments, and the gray dashed line is obtained from the exact solution of the system. Close to the transition the standard deviation increases due to the statistical errors that exchange the depth of the two minima, as we see from the inset histograms. (d) Same as (b), for an anharmonic potential with $\epsilon = 1.5$. (e) Similarly to (c), the optimal value $w_{\rm opt}$ as a function of $\omega$ obtained from $10^4$ trajectories, for different values of $\epsilon$. The gray dashed line corresponds to the harmonic case, i.e. $\epsilon = 0$. Notice that, the larger $\epsilon$, the sooner with respect to $\omega$ the transition happens.}
    \label{fig:figure4}
\end{figure*}

So far, we have studied the case of an harmonically bounded particle starting from the exact expression of $P_{\rm exp}$ without relying on trajectory estimations. Now, we consider the more realistic case in which we only have access to a few, and possibly short-time, trajectories. From these, we extract: (i) the mean, $\mu_{\rm exp}$, and the variance, $\sigma^2_{\rm exp}$, to obtain the maximum entropy ansatz of $P_{\rm exp}$; (ii) the initial conditions in the $x$-space, $\mu_{x_0}$ and $\sigma_{x_0}$; (iii) the \textit{observed} steady state. Then, we obtain the optimal parameters from Eq.~\eqref{minM}.

We first test this trajectory-dependent approach using simulated trajectories for the harmonic case, comparing its results with the analytical ones obtained above. In Fig.~\ref{fig:figure4}b-c, we show that the method generally leads to accurate results. However, close to the transition, the number of trajectories plays a crucial role. Indeed, a small sample size induces statistical errors in the estimates of $\mu_{\rm exp}$ and $\sigma^2_{\rm exp}$, as well as initial conditions and steady states. These errors, in turn, might lead to an inaccurate estimation of the deepest minimum. In the insets of Fig.~\ref{fig:figure4}c we show that, at fixed $\omega \approx \omega_c$, for a small number of trajectories one can end up with a value of $w_{\rm opt}$ corresponding to either of the two minima.

Finally, we apply our method to simulated trajectories generated with an anharmonic force, i.e., $F = \omega^2 x + \epsilon x^3$ (see Fig.~\ref{fig:figure4}a). In this case, the maximum entropy ansatz for $P_{\rm exp}$ is not exact, hence we do not have an exact baseline. Remarkably, the information-driven transition is still present, albeit slightly shifted with respect to the case of an harmonic potential. The number of trajectories plays the same role as before, i.e., it introduces uncertainty close to the transition as it decreases. This result highlights that the transition is a robust feature and has to be considered when building projections that capture the maximum amount of information of a (relatively small) set of experimental trajectories. Indeed, without knowing the parameters of the underlying system, the presence of an abrupt transition might lead to markedly different behaviors of the effective model.

Summarizing, we introduced a method to build information-preserving projections of (possibly unknown) underdamped dynamics. We unraveled an uncharted discontinuity in the optimal parameter space triggered by the minimization of the information loss. This may be interpreted as an information-driven discontinuous transition that induces abrupt changes in the effective model. Our results pose fundamental challenges to the ambition of inferring underlying parameters from effective low-dimensional models, as the appearance of this transition in paradigmatic systems translates into an alarming warning signal for more general cases.

We remark that our approach did not consider thermodynamic features. Non-equilibrium thermodynamics suffers from coarse-graining procedures \cite{esposito2012stochastic,busiello2019entropyA,busiello2019entropyB}, and building projections that preserve the underlying thermodynamics represents a completely different and far-reaching task. A fascinating idea will be to simultaneously optimize dynamics and thermodynamics in a Pareto-like multi-optimization problem \cite{seoane2015phase}.

A more immediate extension of our method would be to perturbatively include higher moments in the estimation of $P_{\rm exp}$ from the experimental data to improve the Gaussian ansatz proposed here. In principle, if a large number of trajectories is accessible, one can directly estimate the full marginal distribution numerically. Moreover, more general classes of effective models should be explored, with a particular attention to understating if and how the corresponding optimal model improves upon a classical overdamped limit.

Ultimately, we believe that our work sheds a light on fundamental properties of effective representations of complex dynamics. Indeed, emerging singularities in low-dimensional models, while crucial in shaping their behavior, might be a sheer consequence of the employed projection method, without reflecting any property of the original system.

\begin{acknowledgments}
A.M. is supported by ``Excellence Project 2018'' of the Cariparo foundation.
\end{acknowledgments}

\pagebreak

\newpage
\widetext
\pagebreak

\setcounter{equation}{0}
\setcounter{figure}{0}
\setcounter{table}{0}
\setcounter{page}{1}
\setcounter{section}{0}
\setcounter{subsection}{0}
\makeatletter
\renewcommand{\theequation}{S\arabic{equation}}
\renewcommand{\thefigure}{S\arabic{figure}}
\renewcommand{\thesection}{S\Roman{section}} 


\begin{center}\Large{Supplemental Material: ``Information-driven transitions in projections of underdamped dynamics'' }\end{center}

\section*{A. Long-time limit and constraint-free optimization}
As in the main text, we compute the probability distribution $P_{\rm exp}(\vec{x}, t)$ from a set of observed trajectories in the $\vec x$-space with duration $t_{\rm exp}$. The Gaussian ansatz prescribes that this probability distribution is fully determined by its mean $\mu_{\rm exp}(t)$ and variance $\sigma^2_{\rm exp}(t)$. To build an information projection of the observed dynamics, we seek the optimal model parameters
\begin{equation}
    {\vec \theta_{\rm opt}} = \underset{{\vec \theta_p}}{\rm argmin} \int_0^{t_{\rm exp}} dt~\mathcal{M}(P_{\rm exp}, P_{\rm OU})
    \label{SM:minM}
\end{equation}
where $\mathcal{M}$ is an information metric. Here, ${\vec \theta_p} := \{ w_p$, $\vec{\mu}_p$, $\hat{\Delta}_p \}$ define the Ornstein-Uhlenbeck process
\begin{equation}
    \dot{\vec{x}}(t) = - w_p \left(\vec{x}(t) - \vec{\mu}_p \right) + \sqrt{2} \hat{\Delta}_p \vec{\xi}_p(t)
    \label{SM:OU}
\end{equation}
whose solution is $P_{\rm OU}(\vec{x}, t)$. In general, as $t_{\rm exp} \to +\infty$, the integral in Eq.~\eqref{SM:minM} might diverge since $\mathcal{M}(P_{\rm exp}, P_{\rm OU})$ may tend to a constant non-zero value in the long-time limit. In fact, $\mathcal{M}(P_{\rm exp}, P_{\rm OU})$ is zero if and only if $P_{\rm exp} = P_{\rm OU}$.

To avoid such divergence, as stated in the main text, we impose that the stationary limit $t\to+\infty$ of $P_{\rm OU}(\vec{x}, t)$ is the same of $P_{\rm exp}(\vec{x}, t)$. In this way, we are guaranteed that $\lim_{t\to\infty}\mathcal{M}(P_{\rm exp}, P_{\rm OU}) = 0$. This choice amounts to add a Lagrange multiplier to the minimization in Eq.~\eqref{SM:minM}. Notably, this constraint effectively reduces the number of free parameters ${\vec \theta_p}$. For instance, in the $1D$ harmonic case considered in the main text, the stationary limit of $P_{\rm exp}$ is defined by $\mu^{\rm stat}_{\rm exp} = 0$ and $\sigma^{\rm stat}_{\rm exp} = \sqrt{k_B T}/\omega$. On the flip side, the steady state of the Ornstein-Uhlenbeck process determines $x_{\rm OU}^{\rm stat} = \mu_p$ and $\sigma_{\rm OU}^{\rm stat} = \Delta_p/w_p$. As a consequence, $\mu_p = 0$, $\Delta_p/\omega_p = \sigma^{\rm stat}_{\rm exp}$. As we optimize only over $\omega_p$, then $\Delta_p$ is automatically determined by the constraint.

Notice that in the multi-dimensional case, the constraint above will reduce the number of model parameters, but they are generally more than one.

At any rate, we can relax this condition on the steady states of $P_{\rm exp}$ and $P_{\rm OU}$, by employing the integral mean in Eq.~\eqref{SM:minM} in order to avoid divergences in the long-time limit:
\begin{equation}
    {\vec \theta_{\rm opt}} = \underset{{\vec \theta_p}}{\rm argmin} \frac{1}{t_{\rm exp}} \int_0^{t_{\rm exp}} dt~\mathcal{M}(P_{\rm exp}, P_{\rm OU}).
    \label{SM:minM}
\end{equation}

\section*{B. Analytic solution in the harmonic case}
We consider the underdamped dynamics of a particle in a one-dimensional harmonic potential, described by the set of Langevin equations
\begin{align}
    \begin{cases}
    \dot{x}(t) = v(t) \\
    \dot{v}(t) = -\omega^2 x(t) -\gamma v(t) + \sqrt{2 k_B T \gamma} \xi(t)
    \end{cases}
\end{align}
where $\xi(t)$ is a Gaussian white noise, and $m$ is set to $1$. These Langevin equations correspond to the Kramers equation
\begin{align}
        \partial_ t p(x, v, t) + v \partial_x p(x, v, t) = \partial_v[(\omega^2 x + \gamma v) p(x, v, t)] + v_{\rm th}^2 \gamma \partial_v^2 p(x, v, t),
        \label{kramersHBP}
    \end{align}
with $v_{\rm th}^2 = k_B T$. We can solve this equation for the propagator $p(x, v, t | x_0, v_0, 0) \sim \mathcal{N}(\vb{M}, S)$, where
\begin{gather*}
    \vb M = e^{-\Gamma t} \begin{pmatrix}x_0 \\ v_0\end{pmatrix} \\
    S_{xx} = \frac{\gamma  v_{\rm th}^2}{(\lambda_1-\lambda_2)^2} \left(\frac{\lambda_1+\lambda_2}{\lambda_1 \lambda_2}+\frac{4 \left(e^{t (-\lambda_1-\lambda_2)}-1\right)}{\lambda_1+\lambda_2}-\frac{e^{-2 \lambda_1 t}}{\lambda_1}-\frac{e^{-2 \lambda_2 t}}{\lambda_2}\right) \\
    S_{vv} = \frac{\gamma  v_{\rm th}^2}{(\lambda_1-\lambda_2)^2} \left(\lambda_1+\lambda_2+\frac{4\lambda_1\lambda_2 \left(e^{t (-\lambda_1-\lambda_2)}-1\right)}{\lambda_1+\lambda_2}-e^{-2 \lambda_1 t}\lambda_1-e^{-2 \lambda_2 t}\lambda_2\right) \\
    S_{xv} = S_{vx} = \frac{\gamma  v_{\rm th}^2}{(\lambda_1-\lambda_2)^2}\left(e^{-\lambda_1 t} + e^{-\lambda_2 t}\right)^2
\end{gather*}
and
\begin{align*}
    \Gamma = \begin{pmatrix}
    0 & -1 \\
    \omega^2 & \gamma \\
    \end{pmatrix}, \quad\quad \lambda_{1, 2} = \frac{\gamma \pm \sqrt{\gamma^2-4\omega^2}}{2}.
\end{align*}
We further consider that the initial conditions $(x_0, v_0)$ are described by two independent Gaussian distributions $p(x_0) \sim \mathcal{N}(\mu_{x_0}, \sigma_{x_0}^2)$ and $p(v_0) \sim \mathcal{N}(\mu_{v_0}, \sigma_{v_0}^2)$, so that
\begin{align}
    p(x, v, t) \sim \mathcal{N}\left(e^{-\Gamma t} \begin{pmatrix}\mu_{x_0} \\ \mu_{v_0}\end{pmatrix}, S + e^{-\Gamma t} \begin{pmatrix}\sigma_{x_0}^2 & 0 \\ 0 & \sigma_{v_0}^2\end{pmatrix}\left(e^{-\Gamma t}\right)^T\right)
\end{align}
is the solution to the Kramers equation we are seeking.

As in the main text, we are not interested in the complete description of the system, but rather in the position space alone. Since we know the full analytical solution of the system, this assumption amounts to computing the marginal probability distribution over the $x$-space $p_{\rm har}(x, t) = \int dv \, p(x, v, t)$. Clearly, this is still a Gaussian distribution, with mean and variance
\begin{align*}
    \mu_{\rm har}(t) & = e^{-\frac{\gamma t}{2}} \left[\frac{\gamma\mu_{x_0}+2 \mu_{v_0} \sinh\left(\frac{\lambda t}{2}\right)}{\lambda}+\mu_{x_0} \cosh\left(\frac{\lambda t}{2}\right)\right] \\
    \sigma^2_{\rm har}(t) & = \frac{e^{-\gamma t}}{\omega^2\lambda^2} \biggl[\gamma \sigma_{x_0}^2 \omega ^2\lambda \sinh \left(\lambda t\right) + \cosh \left(\lambda t\right) \left(\omega^2 \left(\sigma_{x_0}^2(\gamma^2 - 2\omega^2)+2 \sigma_{v_0}^2\right)-\gamma ^2 v_{\rm th}^2\right) + \\
    & \quad\quad\quad\quad + v_{\rm th}^2\lambda\left[\lambda e^{\gamma t}-\gamma\sinh(\lambda  t)\right] - 2\omega ^2\left(\sigma_{v_0}^2+\sigma_{x_0}^2 \omega ^2-2 v_{\rm th}^2\right)\biggl]
\end{align*}
where $\lambda = \sqrt{\gamma ^2-4 \omega ^2}$. Hence, employing the Gaussian ansatz, $\mu_{\rm exp}(t) = \mu_{\rm har}(t)$ and $\sigma_{\rm exp}(t) = \sigma_{\rm har}(t)$ in this example. Since we are studying the system until stationarity, $t_{\rm exp} \to +\infty$.

Here, we also report for completeness the expression of $\mu_{\rm OU}(t)$ and $\sigma_{\rm OU}(t)$ for the $1D$ case below. The generalization to the multi-dimensional case is straightforward.
\begin{align*}
    \mu_{\rm OU}(t) & = \mu_p (1 - e^{-w_p t}) + \mu_{x_0}e^{-w_p t} \\
    \sigma^2_{\rm OU}(t) & = \frac{\Delta_p^2}{w_p} (1 - e^{-2 w_p t}) + \sigma_{x_0}^2 e^{-2 w_p t}
\end{align*}

\section*{C. Results for different information and non-information metrics}
In the main text, we choose as information metric $\mathcal{M}$ the symmetrized Kullback-Leibler divergence. That is, for two $1D$ Gaussian distributions $P_{\rm OU}$ and $P_{\rm exp}$,
\begin{align}
    \mathcal{F}_\mathcal{M} & = \int_{0}^\infty \frac{1}{2} \left[D_{\rm KL}(P_{\rm exp} || P_{\rm OU}) + D_{\rm KL}(P_{\rm OU} || P_{\rm exp})\right]dt \nonumber \\
    & = \int_{0}^\infty \left[\frac{\sigma^2_{\rm OU}(t) + (\mu_{\rm OU}(t) - \mu_{\rm exp}(t))^2}{4 \sigma^2_{\rm exp}(t)} + \frac{\sigma^2_{\rm exp}(t) + (\mu_{\rm exp}(t) - \mu_{\rm OU}(t))^2}{4 \sigma^2_{\rm OU}(t)}\right] dt
\end{align}
which is the quantity that we minimize in the main text. 

Here, we show the appearance of the same phenomenology when considering different information distances. Clearly, each distance has a different information-geometric meaning, hence the exact transition line $\omega_c(\gamma)$ will change, but the discontinuous transition is always present - even if the functional forms of the metrics are vastly different. We employ
\begin{gather}
    H^2(P_{\rm exp}(x, t), P_{\rm OU}(x, t)) = 1 - \sqrt{2\frac{\sigma_{\rm exp}\sigma_{\rm OU}}{\sigma_{\rm exp}^2 + \sigma_{\rm OU}^2}}e^{-\frac{1}{4}\frac{(\mu_{\rm exp} - \mu_{\rm OU})^2}{\sigma_{\rm exp}^2 + \sigma_{\rm OU}^2}} \\
    G(P_{\rm exp}(x, t), P_{\rm OU}(x, t)) = 2\sqrt{2} \tanh^{-1}\left[\sqrt{\frac{(\mu_{\rm exp} - \mu_{\rm OU})^2 + 2(\sigma_{\rm exp} - \sigma_{\rm OU})^2}{(\mu_{\rm exp} - \mu_{\rm OU})^2 + 2(\sigma_{\rm exp} + \sigma_{\rm OU})^2}}\right] \\
    C_\alpha(P_{\rm exp}(x, t) || P_{\rm OU}(x, t)) = \frac{\alpha(1-\alpha)}{2}\frac{(\mu_{\rm exp} - \mu_{\rm OU})^2}{(1-\alpha)\sigma_{\rm exp}^2 + \alpha\sigma_{\rm OU}^2} + \frac{1}{2}\log\left[\frac{(1-\alpha)\sigma_{\rm exp}^2 + \alpha\sigma_{\rm OU}^2}{\sigma_{\rm exp}^{2(1-\alpha)} + \sigma_{\rm OU}^{2\alpha}}\right] \\
    W_2^2(p_{\rm exp}(x, t), P_{\rm OU}(x, t)) = \sqrt{(\mu_{\rm exp} - \mu_{\rm OU})^2} + \sigma_{\rm exp}^2 + \sigma_{\rm OU}^2 + \sqrt{\sigma_{\rm exp}^2 \sigma_{\rm OU}^2}
\end{gather}
which are, respectively, the Hellinger distance, the geodetic distance, the Chernoff-alpha divergence and the Wasserstein distance between two $1D$ Gaussian distributions. 

\begin{figure}
    \centering
    \includegraphics[width = 0.9 \columnwidth]{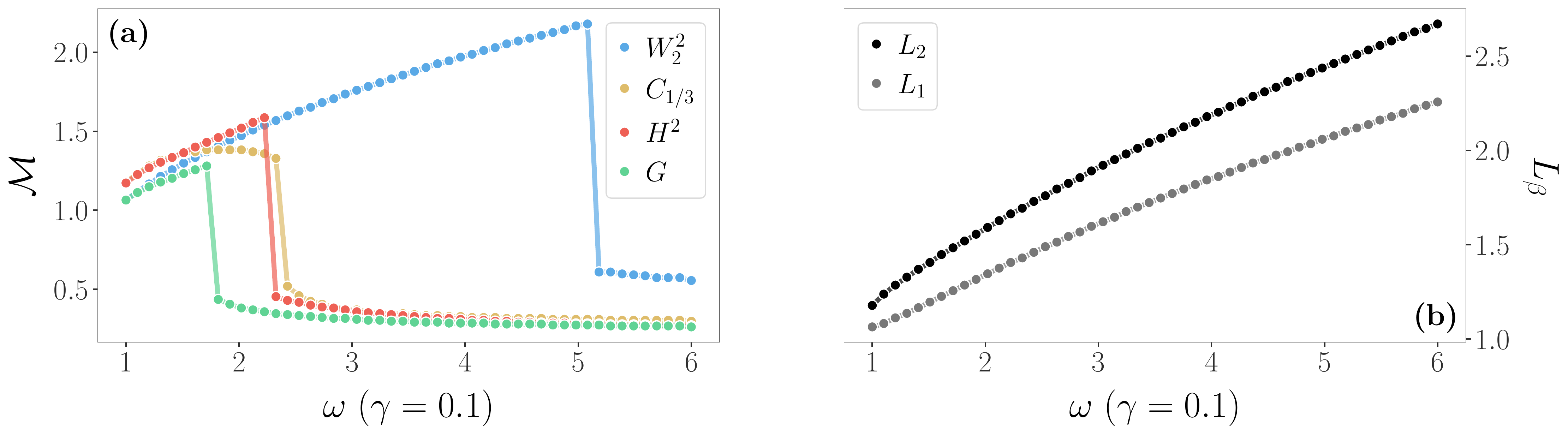}
    \caption{$w_{\rm opt}$ for different metrics as a function of $\omega$, with $\gamma = 0.1$. (a) Different colors correspond to different choices of the information metric. They all exhibit a discontinuous transition, even if for different $\omega_c(\gamma)$. (b) Non-information metrics do not present transitions, and $w_{\rm opt}$ monotonously increase with $\omega$. The arbitrary choice of $\gamma$ helps visualizing the results.}
    \label{figS1}
\end{figure}

In Fig.~\ref{figS1}a we show the discontinuous transition for a specific value of $\gamma$, as a function of $\omega$. This result strongly suggests that the transition is an intrinsic feature stemming from the minimization of the information loss, and not just a byproduct of our specific choice of the metric.

Moreover, in Fig.~\ref{figS1}b we explore the same region of the parameter space minimizing a non-information metric. No transition is present in this case, and $w_{\rm opt}$ monotonously increases with $\gamma$. Specifically, we use
\begin{eqnarray}
L_\beta(\mu_{\rm exp},\sigma^2_{\rm exp},\mu_{\rm OU},\sigma^2_{\rm OU}) = \left|\mu_{\rm exp} - \mu_{\rm OU}\right|^\beta + \left|\sigma^2_{\rm exp} - \sigma^2_{\rm OU}\right|^\beta
\end{eqnarray}
Notice that the non-information metrics are characterized by the fact that they are not distances in the probability space, hence depending solely on the mean and variance of $P_{\rm exp}$ and $P_{\rm OU}$.

\section*{D. Effect of the initial conditions}
\begin{figure}
    \centering
    \includegraphics[width = 0.9 \columnwidth]{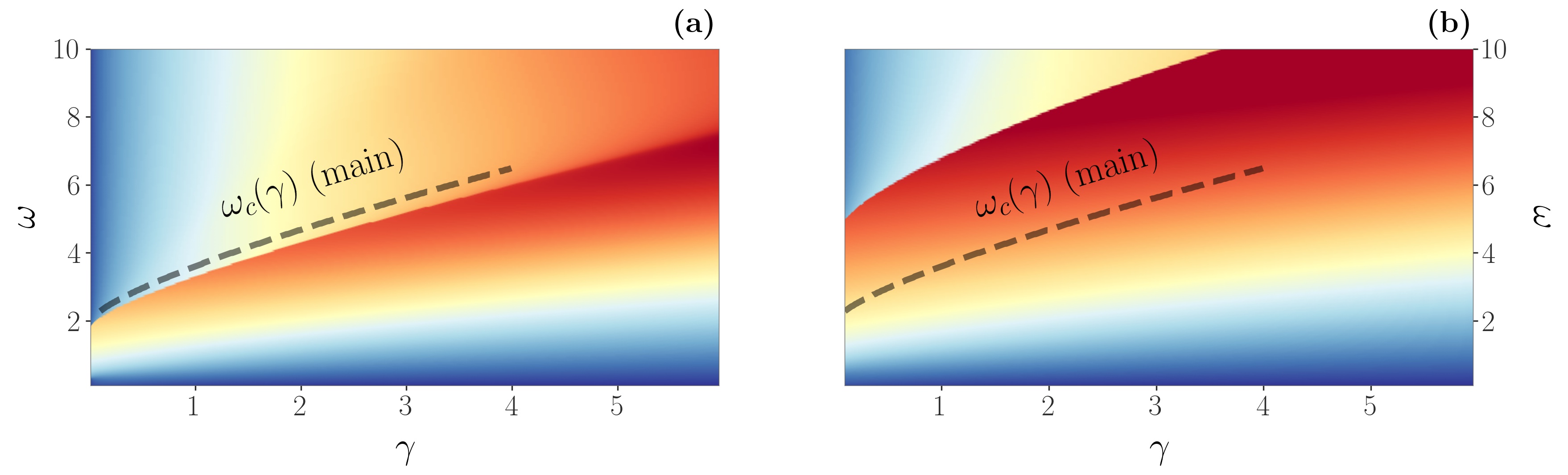}
    \caption{Contour plot of $w_{\rm opt}$ as a function of $\omega$ and $\gamma$ for (a) $v_{\rm th}^2 = 0.75$, $\mu_{x_0} = 1.25$, $\sigma_{x_0} = 1$, $\mu_{v_0} = 0$, $\sigma_{v_0}^2 = 0.5$, and (b) $\mu_{x_0} = 1$, $\sigma_{x_0} = 0.5$, $\mu_{v_0} = 0$, and $\sigma_{v_0}^2 = 1$. The dashed lines indicate the transition line $\omega_c(\gamma)$ for the initial conditions of the main text.}
    \label{figS2}
\end{figure}

In Fig.~\ref{figS2} we show that as we change the initial conditions, the qualitative picture stays the same, while the transition line changes in the $(\omega,\gamma)$-space. We also remark that a variation in $\sigma_{x_0}^2$ (Fig.~\ref{figS2}b) has a greater impact than a variation in all the other initial conditions (Fig.~\ref{figS2}a). Notice that we also changed $v_th^2$ in this latter panel only to explore different parameters from the ones presented in the main text.

\section*{E. Results for constant variance in the harmonic case}
As discussed in the main text, tuning the initial conditions so that the variance is constant at all times, and equal to the stationary variance, the transition disappears. In this case we can solve the problem analytically. Indeed, setting $\sigma_{x_0}^2 = v_{\rm th}^2/\omega^2$ and $\sigma_{v_0}^2 = v_{\rm th}^2 = k_B T$, that $\sigma_{\rm har}^2(t) = \sigma_{\rm OU}^2(t) = \sigma_{x_0}^2$. Hence, the Kullback-Leibler divergence depends only on the mean and greatly simplifies. We have to minimize
\begin{align*}
    \mathcal{F}_\mathcal{M}(w_p) & = \int_{0}^\infty \frac{(\mu_{\rm OU}(t) - \mu_{\rm har}(t))^2}{2v^2_{\rm th}/\omega^2} dt \\
    & = \frac{\omega^2}{2 v_{\rm th}^2 \lambda^2}\int_0^{\infty} e^{-t \left(\gamma +2 w_p^2\right)} \left[\mu_{x_0} \lambda e^{\frac{\gamma t}{2}}-e^{t w_p^2}f(t; \gamma, \omega, \mu_{x_0}, \mu_{v_0}) \right]^2 \\
    & = \frac{1}{4v_{\rm th}^2}\left[\frac{\mu_{x_0}^2\omega^2 + (\mu_{v_0} + \gamma\mu_{x_0})^2}{\gamma} + \frac{\mu_{x_0}^2\omega^2}{w_p^2} - \frac{4 \mu_{x_0}\omega^2[\mu_{v_0} + \mu_{x_0}(\gamma + w_p^2)]}{\omega^2 + \gamma w_p^2 + w_p^4}\right].
\end{align*}
where $\lambda = \sqrt{\gamma ^2-4 \omega^2}$ and $f(t; \gamma, \omega, \mu_{x_0}, \mu_{v_0}) = (\gamma \mu_{x_0}+2 \mu_{v_0}) \sinh \frac{\lambda t}{2} +\mu_{x_0} \lambda \cosh\frac{\lambda t}{2}$.

Hence, we have to solve
\begin{align*}
     0 & = \frac{4 \mu_{v_0} w_p^4 \left(\gamma +2 w_p^2\right)-\mu_{x_0} \left(2 \omega ^2 w_p^2 \left(\gamma +3 w_p^2\right)-3 w_p^4 \left(\gamma +w_p^2\right)^2+\omega^4\right)}{w_p^3 \left(\gamma  w_p^2+\omega^2+w_p^4\right)^2} \\
     & = 4 \mu_{v_0} w_p^4 \left(\gamma +2 w_p^2\right)-\mu_{x_0} \left(2 \omega ^2 w_p^2 \left(\gamma +3 w_p^2\right)-3 w_p^4 \left(\gamma +w_p^2\right)^2+\omega^4\right)
\end{align*}
which has a solution that is always positive and analytical, although particularly cumbersome and not reported here. Notably, when $\mu_{v_0} = 0$, we only need to find the positive and real solution of the equation
\begin{align*}
    \omega^4 + 2\gamma\omega^2w_p^2-3(\gamma^2-2\omega^2)w_p^4-6\gamma w_p^6-3w_p^8 = 0
\end{align*}
which does not depend on $\mu_{x_0}$. We find
\begin{align*}
    w_{\rm opt} = \sqrt{\frac{\sqrt{-\frac{8 \sqrt{3} \gamma  \omega ^2}{\sqrt{A(\gamma, \omega)+B(\gamma, \omega)}}-A(\gamma, \omega)+2 B(\gamma, \omega)}+\sqrt{A(\gamma, \omega)+B(\gamma, \omega)}}{2 \sqrt{3}}-\frac{\gamma}{2}}
\end{align*}
where
\begin{align*}
    A(\gamma, \omega) & = \frac{\gamma^4}{\sqrt[3]{8 \omega ^3 \sqrt{16 \omega ^6-\gamma ^6}+\gamma ^6-32 \omega ^6}} + \sqrt[3]{8 \omega ^3 \sqrt{16 \omega ^6-\gamma ^6}+\gamma ^6-32 \omega ^6} \\
    B(\gamma, \omega) & = \gamma^2 + 4\omega^4.
\end{align*}
Remarkably, if we expand this solution for $\gamma \to +\infty$ we find
\begin{equation*}
    w_{\rm opt} = \frac{\omega}{\sqrt{\gamma}} - \frac{3\omega^5}{8}\frac{1}{\gamma^{9/2}} + \mathcal{O}\left(\frac{1}{\gamma^{11/2}}\right)
\end{equation*}
which is the correction to the overdamped solution. Instead, when $\gamma \to 0$, we have
\begin{equation*}
    w_{\rm opt} = \left(1+\frac{2}{\sqrt{3}}\right)^{1/4}\sqrt{\omega} - \frac{(3+2\sqrt{3})^{1/4}}{2\sqrt{6\omega}}\gamma + \mathcal{O}\left(\gamma^{4/3}\right)
\end{equation*}
so the behavior in the small-$\gamma$ regime is drastically different. In particular, the zero-th order approximation of $w_{\rm opt}$ does not depend on $\gamma$ anymore. Clearly, in this case there is no discontinuous transition, since $\mathcal{F}_\mathcal{M}(w_p)$ has always a unique minimum.
\end{document}